\documentstyle[12pt]{article}
\advance\voffset by -0.8in
\advance\hoffset by -0.55in
\newcommand{\Bbb}{\bf}
\newcommand{\Hil}{{\cal H}}
\newcommand{\ce}{{\Bbb C}}
\newcommand{\ze}{{\Bbb Z}}
\newcommand{\re}{{\Bbb R}}
\newcommand{\reg}[1]{(\ref{#1})}
\newcommand{\implies}{{\Rightarrow}}
\begin{document}
\title{Continuous Symmetries of Lattice Conformal Field Theories and their
$Z\!\!\!Z_2$-Orbifolds}
\author{\underline{P.S. Montague} \\ Department of Applied Mathematics and
Theoretical Physics\\ University of Cambridge\\
Silver Street\\ Cambridge CB3 9EW}
\maketitle
\begin{abstract}
Following on from a general observation in an earlier paper
\cite{PSMorb}, we consider the
continuous symmetries of a certain class of conformal field theories
constructed
from lattices and their reflection-twisted
orbifolds. It is shown that the naive expectation that the only such (inner)
symmetries are
generated by the modes of the vertex operators corresponding to the
states of unit conformal weight obtains, and a criterion for this
expectation to hold in general is proposed.
\end{abstract}
\vfill
\eject
\section{Introduction}
The identification of the automorphism group of a conformal field theory, while
being
clearly of import in its own right as a guide towards an understanding of the
general structure
of the theory, is an essential tool in the classification program.
It has been demonstrated in \cite{DGMtrialsumm,DGMtriality} that the
FKS lattice
theories
and their reflection-twisted orbifolds \cite{DGMtwisted}
(see section \ref{definitions} for definitions)
form a crucial element in the classification
of self-dual conformal field theories. Such a classification needs to be
performed separately from any
mainstream approach involving consideration of fusion rules
\cite{Verlinde},
since the self-dual theories are
trivial from this point of view, and physically is relevant in heterotic string
theory \cite{Schell:Venkov}.
Steps towards this classification have been accomplished from two opposite
approaches.
On the one hand, Schellekens has restricted the possible algebras which
correspond to the weight one
states in the theories at central charge 24
\cite{SchellComplete,Schell:seventy},
while, on the other hand, constructions of
theories which exhibit these algebras have been accomplished \cite{DGMtwisted}.
The only theories
constructed so far are the FKS lattice theories and their reflection-twisted
orbifolds. It is generally
believed that the remaining theories are orbifolds of some form (either of the
FKS theories
themselves, or of orbifolds of these theories), and investigations are
proceeding
along these lines (see {\em e.g.} \cite{PSMorb}). As a result, the
identification of the automorphism groups of the FKS
theories and their only known consistent orbifolds (the reflection-twisted
theories) are of import
in as much as enabling a full classification of all orbifolds which may be
obtained.

Some comments on the discrete part of the automorphism group of these theories
are contained
in \cite{PSMorb}. In this letter, we discuss the continuous symmetries of the
theories.

We begin in section \ref{definitions} by briefly describing the FKS
lattice conformal field theories $\Hil(\Lambda)$ and their reflection-twisted
orbifolds $\widetilde\Hil(\Lambda)$ ($\Lambda$ an even lattice).
This is simply a summary of the relevant parts of previous
work \cite{DGMtwisted}. We then summarize (and slightly refine) in
section \ref{continuous}
the comments made in \cite{PSMorb} on
continuous symmetries in general, before proceeding to apply our considerations
in sections \ref{straight} and \ref{twisted} to $\Hil(\Lambda)$ and
$\widetilde\Hil(\Lambda)$ respectively.

Our conclusions are presented in section \ref{conclusions}.
\section{The conformal field theories $\Hil(\Lambda)$ and
$\widetilde\Hil(\Lambda)$}
\label{definitions}
Throughout this paper, the term conformal field theory shall be taken
to mean bosonic chiral meromorphic conformal
field theory.
$\Hil$ is a (bosonic chiral meromorphic) conformal
field theory if \cite{PGmer} $\Hil$ is
a Hilbert space equipped with a set of linear {\em vertex} operators
$V(\psi,z):\Hil\rightarrow\Hil$ ($\psi\in\Hil$, $z\in\ce$) such that
for $\psi,\phi\in\Hil$
\begin{equation}
V(\psi,z)V(\phi,w)=V(\phi,w)V(\psi,z)\,,
\end{equation}
in the sense of analytic continuation in the complex variables $z$ and
$w$ of the meromorphic matrix elements we may obtain from either side (the
left hand side, for example, being strictly only defined for
$|z|>|w|$).
It may be shown (and this is often taken as an axiom) that
\begin{equation}
V(\psi,z)V(\phi,w)=V\left(V(\psi,z-w)\phi,w\right)\,,
\label{ope1}
\end{equation}
again with suitable analytic continuation. This is the so-called {\em
  operator product expansion} or duality relation.
There are also preferred states $|0\rangle,\psi_L\in\Hil$ such
that
\begin{equation}
V(\psi_L,z)\equiv\sum_nL_nz^{-n-2}\,,
\end{equation}
where
\begin{equation}
[L_m,L_n]=(m-n)L_{m+n}+{c\over 12}m(m^2-1)\delta_{m,-n}
\end{equation}
for some {\em central charge} $c$, and $|0\rangle$ is an SU(1,1)
invariant state (annihilated by $L_{\pm 1},L_0$) such that
\begin{equation}
\label{creater}
V(\psi,z)|0\rangle=e^{zL_{-1}}\psi\,.
\end{equation}
The space may be decomposed into eigenstates of $L_0$, the eigenvalues
being known as the {\em conformal weights}.
In general, for $\psi$ of conformal weight $h$, we write
\begin{equation}
V(\psi,z)\equiv\sum_nV(\psi)_nz^{-n-h}\,,
\end{equation}
and we find that the action of $V_n(\psi)$ shifts the conformal weight
of a state by $n$.
Fuller details and consequences of this set of axioms may be found in
\cite{DGMtriality}.

We now define a specific class of conformal field theories which will
be of relevance in this paper.

Suppose we are given a $d$-dimensional even (Euclidean) lattice
$\Lambda$. We shall define a conformal field theory denoted
$\Hil(\Lambda)$, the FKS lattice theory referred to in the
introduction.

We introduce a set of bosonic creation and annihilation operators
$a_n^i$, $n\in\ze$, $1\leq i\leq d$, such that
\begin{eqnarray}
[a_m^i,a_n^j]&=&m\delta^{ij}\delta_{m,-n}\nonumber\\
{a^i_m}^\dagger&=&a^i_{-m}\,.
\end{eqnarray}
Let us denote $a_0^i$ by $p^i$ (the momentum operator)
and also introduce an operator
$q^i(={q^i}^\dagger)$ such that $[q^i,p^j]=i\delta^{ij}$. The Hilbert
space of $\Hil(\Lambda)$ is composed of all linear combinations of
states of the form
\begin{equation}
\label{stater}
\prod_{a=1}^Na_{-n_a}^{i_a}|\lambda\rangle\,,
\end{equation}
where $n_a\in\ze_+$ and $\lambda\in\Lambda$ with
$a_n^i|\lambda\rangle=0$ for $n>0$ and
$p^i|\lambda\rangle=\lambda^i|\lambda\rangle$.
Set
\begin{equation}
X^i(z)=q-ip\log z+i\sum_{n\neq 0}{a^i_n\over n}z^{-n}\,,
\end{equation}
and then
\begin{equation}
V\left(\prod_{a=1}^Na^{i_a}_{-n_a}|\lambda\rangle,z\right)=
:\prod_{a=1}^{N}{1\over (n_a-1)!}{d^{n_a-1}X^{i_a}(z)\over
dz^{n_a-1}}e^{i\lambda\cdot X(z)}:\sigma_\lambda\,,
\label{vop}
\end{equation}
where the $\sigma_\lambda$ are a set of cocycle operators satisfying
\begin{equation}
\label{gamma}
\sigma_\lambda\sigma_\mu=(-1)^{\lambda\cdot\mu}\sigma_\mu\sigma_\lambda
\end{equation}
for $\lambda$, $\mu\in\Lambda$,
and we use the usual normal ordering convention on the oscillators.
This is sufficient (see {\em e.g.} \cite{DGMtwisted})
to make $\Hil(\Lambda)$ into a consistent conformal
field theory (with central charge
$d$ and $\psi_L={1\over 2}a_{-1}\cdot a_{-1}|0\rangle$). The conformal
weight of the state \reg{stater} is simply $\sum_{a=1}^Nn_a+
{1\over 2}\lambda^2$. Physically,
this theory represents a bosonic string propagating on the torus
$\re^d/\Lambda$.

Now, the lattice clearly admits a reflection symmetry $\lambda\mapsto
-\lambda$, and this trivially lifts to an involution $\theta$ of
$\Hil(\Lambda)$ ($\theta a^i_n\theta^{-1}=-a^i_n$, $\theta|\lambda\rangle=
|-\lambda\rangle$). Let
$\Hil(\Lambda)_+$ be the sub-conformal field theory on which
$\theta=1$.

Suppose now that the dimension $d$ of the lattice is a multiple of 8.
We construct \cite{DGMtriality,FHL} a meromorphic representation
$\Hil_T(\Lambda)_+$ of
$\Hil(\Lambda)_+$ as follows. The Hilbert space is built up from a
ground state of conformal dimension ${d\over 16}$ (which forms an
irreducible representation space for a set of gamma matrices
$\gamma_\lambda$, $\lambda\in\Lambda$, satisfying
$\gamma_\lambda\gamma_\mu=(-1)^{\lambda\cdot\mu}\gamma_\mu\gamma_\lambda$
-- {\em c.f.} \reg{gamma})
by the action of an odd or even number (according as $d$ is an odd or
an even multiple of 8) of
half-integrally graded bosonic creation and annihilation operators
$c^i_r$, $r\in\ze+{1\over 2}$, $1\leq i\leq d$, such that
\begin{eqnarray}
[c_r^i,c_s^j]&=&r\delta^{ij}\delta_{r,-s}\nonumber\\
{c^i_r}^\dagger&=&c^i_{-r}\,,
\end{eqnarray}
($c_r^i$ annihilates the ground state for $r>0$).
Vertex operators $V_T(\psi,z)$, $\psi\in\Hil(\Lambda)_+$ may be defined
\cite{DGMtwisted}
such that they form a representation of $\Hil(\Lambda)_+$, {\em i.e.}
\begin{equation}
V_T(\psi,z)V_T(\phi,w)=V_T\left(V(\psi,z-w)\phi,w\right)\,,
\end{equation}
{\em c.f.} \reg{ope1}.
In \cite{DGMtwisted}, it is then shown how one may define vertex
operators $W(\chi,z)$ and $\overline W(\chi,z)$ corresponding to
states $\chi\in\Hil_T(\Lambda)_+$ such that
\begin{equation}
{\cal V}\left((\psi,\chi),z\right)=\left(
\begin{array}{cc}V(\psi,z) & \overline W(\chi,z) \\
W(\chi,z) & V_T(\psi,z)
\end{array}\right)
\end{equation}
equips the space $\widetilde\Hil(\Lambda)\equiv\Hil(\Lambda)_+\oplus
\Hil_T(\Lambda)_+$ with the structure of a conformal field theory,
provided that $\sqrt 2\Lambda^\ast$ is even (also a necessary
condition \cite{thesis}).
\section{Continuous (inner) automorphisms of meromorphic conformal field
theories}
\label{continuous}
It is a widely believed result that the continuous symmetries of
such a conformal field theory
are fully accounted
for by the well-known Lie algebra defined by the zero modes of the
vertex operators corresponding to the states of conformal weight one.
However, there are clear counter examples to this.

Suppose $\Lambda$ is an even lattice of dimension $d$, as in the previous
section.
We define the rank of $\Lambda$ to be the dimension of the space it
spans within this $d$ dimensional space. Consider the extreme case in
which $\Lambda$ has rank zero, {\em i.e.} set all vectors in the
lattice to be identically zero.
$\Hil(\Lambda)$ is then simply the Hilbert space obtained by acting on
the vacuum with $d$ commuting sets of creation and annihilation
operators
$a_n^i$, and clearly has an O($d$) symmetry group (given by
$a_n^i\mapsto
{R^i}_ja^j_n$).
$\Hil(\Lambda)_+$ also inherits this O($d$) symmetry group,
but there are no states of
conformal weight one (the states $a_{-1}|0\rangle$ having been
projected out).

In general,
let $\theta$ be a continuous symmetry of a conformal
field theory $\Hil$, {\em i.e.}
\begin{equation}
e^{a\theta}V(\psi,z)e^{-a\theta}=V(e^{a\theta}\psi,z)\,,
\end{equation}
for all $\psi\in\Hil$ and $a\in\re$, or
\begin{equation}
[\theta,V(\psi,z)]=V(\theta\psi,z)\,.
\end{equation}
Suppose further that $\theta$ is ``inner", {\em i.e.} it can be written in
terms of the vertex operators
of the theory. Since duality \reg{ope1} allows us to reduce products of vertex
operators,
and any automorphism leaves the Virasoro generators invariant \cite{thesis}
(and hence the
conformal weights), we write $\theta=V(\psi_\theta)_0$ for some state
$\psi_\theta\in\Hil$.
Expand $\psi_\theta=\sum_{n\geq 1}\psi_n$, where $\psi_n$ is of conformal
weight $n$ (we
exclude the case $n=0$ as it simply gives a trivial addition of a constant to
$\theta$).

Now, again using the invariance of the Virasoro generators under the
automorphism, we must
have
$[L_{-1},\theta]=0$, {\em i.e.}
\begin{equation}
\sum_{n\geq 1}(n-1)V_{-1}(\psi_n)=0\,,
\end{equation}
from the relation
\begin{equation}
[L_{-1},V(\psi,z)]=V(L_{-1}\psi,z)={d\over dz}V(\psi,z)\,,
\label{L1CR}
\end{equation}
(equivalent to \reg{creater}).
Suppose that $V_{-1}(\psi)=0\Rightarrow\psi=\lambda|0\rangle$ for some
$\lambda\in\ce$
($\psi_1=V_{-1}(\psi_1)|0\rangle$ and $V(|0\rangle)=1$).
Then we may deduce that $\psi_n=0$ for $n\geq 2$, as required. However, the
assumption is
too strong.
For example, in the theory $\Hil(\Lambda)$, $V_{-1}(a_{-2}|0\rangle)=0$. We
notice that the
relation \reg{L1CR}
gives $V_{-1}(L_{-1}\psi_n)=-(n-1)V_{-1}(\psi_n)$ (which accounts for the
vanishing of $V_{-1}$
in our example), and we absorb this freedom by using it to redefine the
$\psi_n$ to be
quasi-primary, {\em i.e.} annihilated by $L_1$
(the weight one states are automatically quasi-primary, so there is no
problem at $n=1$). Then our new assumption is that, for $\psi$ quasi-primary,
$V_{-1}(\psi)=0\Rightarrow\psi=\lambda|0\rangle$ for some $\lambda\in\ce$.
This gives us the required result, and that it is reasonable will become clear
in the following
sections where we demonstrate that it holds for both
$\Hil(\Lambda)$ and $\widetilde\Hil(\Lambda)$.

Note that another way of phrasing our condition on the conformal field theory
is that,
if $\psi$ is quasi-primary,
$V_0(\psi)$ determines $\psi$ uniquely up to states of conformal weight one
(see appendix
\ref{app2}).
We define a conformal field theory to be {\em deterministic} if it satisfies
this criterion.
Thus, for a deterministic conformal field theory, the continuous inner
automorphisms
are simply generated by the states of conformal weight one.

In the example of $\Hil(\Lambda)_+$ given above, the theory is not
deterministic as there are redundant operators, {\em i.e.} the
momentum operator is identically zero, and so it is trivial to see
that there are (quasi-primary) states with vanishing zero mode.

Conversely, suppose that $\Hil$ is not deterministic, {\em i.e.} suppose we
have a state
$\psi$ (taken to consist only of states of weight at least 2 without loss of
generality) such
that
\begin{equation}
V_{-1}((L_0-1)\psi)\equiv [L_{-1},V_0(\psi)]=0\,.
\label{assume}
\end{equation}
Consider $[V_0(\psi),V(\phi,w)]$ for some state $\phi\in\Hil$.
Applying this to the vacuum, and using \reg{creater}
we obtain
\begin{equation}
V_0(\psi)e^{wL_{-1}}\phi+V(\phi,w)V_0(\psi)|0\rangle\,.
\end{equation}
But, from \reg{assume}, $L_{-1}$ commutes with $V_0(\psi)$, and also
$V_0(\psi)|0\rangle=
0$ ($V_0(\psi)|0\rangle=\gamma|0\rangle$, $\gamma\in\ce$, where
$\gamma=\langle 0|V(z^{-L_0}\psi,z)|0\rangle=\langle
0|e^{zL_{-1}}z^{-L_0}|\psi\rangle=0$).
Then we see that $[V_0(\psi),V(\phi,w)]$
has the same action on the vacuum as $V(V_0(\psi)\phi,w)$, and
so, by the uniqueness theorem \cite{PGmer} ({\em i.e.} that if $W(z)$
has the same action on the vacuum as $V(\rho,z)$ and is local with
respect to the set of vertex operators then $W(z)=V(\rho,z)$), we deduce that
\begin{equation}
[V_0(\psi),V(\phi,w)]=V(V_0(\psi)\phi,w)\,,
\end{equation}
as required. Note however that there is no guarantee that the automorphism will
be non-trivial.
\section{Determinism of the theories $\Hil(\Lambda)$}
\label{straight}
In this and the following section, we demonstrate that the property of
determinism
holds for the known self-dual theories, and thus that it is a sensible
condition to impose.

We consider a state $\psi\in\Hil(\Lambda)$ such that $V_0(\psi)=0$.
(Note that we will assume that the rank of $\Lambda$ is equal to its
dimension, so that we will always be able to distinguish the operators
$a_0^i\equiv p^i$ from zero.)
First decompose $\psi$ according to its momentum, {\em i.e.} write
\begin{equation}
\psi=\sum_{\lambda\in\Lambda}\sum_{s\in I_\lambda}
\prod_{a=1}^{N_\lambda^s}a^{i_{as}^\lambda}_{
-n_{as}^\lambda}|\lambda\rangle\equiv\sum_{\lambda\in\Lambda}\psi_\lambda\,.
\end{equation}
Clearly, we must have $V_0(\psi_\lambda)=0$ for all
$\lambda\in\Lambda$, since $[p,V(\psi_\lambda)]=\lambda
V(\psi_\lambda)$.
Let us initially consider the case $\lambda=0$ for simplicity.

Hence, if we decompose $\psi_0$ into states
\begin{equation}
\psi_0^{n_1\cdots n_d}=\sum_\alpha \rho_1^\alpha\cdots\rho_d^\alpha
|0\rangle\,,
\end{equation}
where $\rho_i^\alpha$ is a set of $n_i$ creation operators with
vectorial index $i$,
we see that we must
have
$V_0(\psi_0^{n_1\cdots n_d})=0$ for all $\{n_i\}$.

First note that, as observed in section \ref{continuous}, we are able to
rewrite our
states by acting with $L_{-1}$. Rather than considering quasi-primary
states, as we did earlier, we show that other conditions may
be imposed. All we have to do is simply specify the form of the state
sufficiently strongly to absorb all of the ambiguity afforded by the
$L_{-1}$ freedom.

For example, suppose that $V_0(\psi)=0$ and that $\psi$ is of some
such specific form, {\em i.e.} it is a linear combination of states
whose $L_{-1}$ descendants span the space and are linearly independent
(note that $L_{-1}\phi=L_{-1}\chi\implies V(L_{-1}(\phi-\chi),z)\equiv
{d\over{dz}}V(\phi-\chi,z)=0\implies \phi-\chi=\mu|0\rangle$ for some
scalar $\mu$). Then we are able to use the $L_{-1}$ freedom to define
a quasi-primary state $\widehat\psi$ such that $V_0(\widehat\psi)=
V_0(\psi)$ $(=0)$. Note that the form of the definition is such that
$\psi=0$ if and only if $\widehat\psi=0$ (we must use the linear
independence referred to above). Now if
$V_0(\widehat\psi)=0\implies
\widehat\psi=\gamma|0\rangle$ for $\widehat\psi$ quasi-primary,
then $\psi=\gamma|0\rangle$ as
required. The converse argument holds, and we see that it is possible
to prove the result using some more convenient means of fixing up the
freedom.

Now, going back to our particular case, we have $V_0(\psi_0^{n_1\cdots
n_d})=0$. Suppose that at least two of the $n_i$ are non-zero ($n_1$
and $n_2$ without loss of generality). Set
$\rho_2^\alpha\cdots\rho_d^\alpha=\phi^\alpha$. Then we have
\begin{equation}
0=V_0\left(\sum_\alpha\rho_1^\alpha\phi^\alpha|0\rangle\right)=
\sum_nV_{-n}(\rho_1^\alpha|0\rangle)V_n(\phi^\alpha|0\rangle)\,,
\end{equation}
(no normal ordering being necessary since the operators commute).
Because of the form of the vertex operators, we then see that we must,
in particular, have either $ V_0(\rho_1^\alpha|0\rangle)=0$ or
$V_0(\phi^\alpha|0\rangle)=0$. We would then have an inductive step,
provided that the restriction chosen to absorb the $L_{-1}$ freedom
carried over from $\psi_0^{n_1\cdots n_d}$ to both
$\rho_1^\alpha|0\rangle$ and $\phi^\alpha|0\rangle$.

It is possible to choose a set of states whose $L_{-1}$ descendants
span the space in such a way, for example by requiring any non-trivial
$\rho_i^\alpha$ to include at least one $a_{-1}^i$. Though the
descendants of such a set are not necessarily linearly independent, a
subset of them can be chosen such that this holds ({\em e.g.} in the
case of two oscillators we choose $\{a_{-1}^ia_{-2n-1}^i\}$).

Thus, we have reduced the problem to one of verifying the result in
the case of a one-dimensional lattice (at least for the zero-momentum-sector
states).

The case of one creation oscillator is trivial.

 Let us consider the case of 2 oscillators.

We choose as a set of basis states $\{a_{-n}a_{-n}|0\rangle\}$
(dropping vectorial indices, as we are restricting to one-dimension).
(N.B. The action of $L_{-1}$ of course keeps us inside the
two-oscillator subspace!) As a simple check that these are sufficient,
consider the generating function for the number of states at each
level. The states themselves give
\begin{equation}
p(x)=\sum_{n=1}^\infty x^{2n}={x^2\over 1-x^2}\,,
\end{equation}
and therefore the states and their $L_{-1}$ descendants give, assuming
linear independence (which is simply checked:

We already have shown that $L_{-1}\phi=L_{-1}\chi\implies\phi-\chi=
\mu|0\rangle$ for some
scalar $\mu$. Therefore, we only have to show that
$a_{-n}a_{-n}|0\rangle$ is not an $L_{-1}$ descendant of the higher
states.

Suppose the converse. Then we can write
\begin{equation}
a_{-n}a_{-n}|0\rangle=L_{-1}\left(\lambda_1a_{-1}a_{-(2n-2)}|0\rangle
+ \cdots + \lambda_{n-1}a_{-(n-1)}a_{-n}|0\rangle\right)\,.
\label{uno}
\end{equation}
Then
\begin{eqnarray}
a_{-n}a_{-n}|0\rangle&=&\lambda_1a_{-2}a_{-(2n-2)}|0\rangle+
(2n-2)\lambda_1a_{-1}a_{-(2n-1)}|0\rangle+\cdots\\
&&+(n-1)\lambda_{n-1}a_{-n}a_{-n}|0\rangle+n\lambda_{n-1}
a_{-(n-1)}a_{-(n+1)}|0\rangle\,.
\label{tre}
\end{eqnarray}
Hence we deduce successively $\lambda_1=0$, $\lambda_2=0$, $\ldots$,
$\lambda_{n-1}=0$, and also $(n-1)\lambda_{n-1}=1$, giving the
required contradiction.
),
\begin{equation}
q(x)={p(x)\over 1-x}={x^2\over (1-x)(1-x^2)}\,.
\end{equation}
Now, the two-oscillator states at level $2n$ are given by
$a_{-1}a_{-(2n-1)}|0\rangle$, $a_{-2}a_{-(2n-2)}|0\rangle$,
$\ldots$, $a_{-n}a_{-n}|0\rangle$, with a similar set for odd levels.
Hence, the generating function is
\begin{equation}
r(x)=\sum_{n=1}^\infty nx^{2n}+\sum_{n=1}^\infty nx^{2n+1}=
q(x)\,,
\end{equation}
as required.

Now suppose that
\begin{equation}
V_0\left(\sum_{n=1}^\infty \lambda_n a_{-n}a_{-n}|0\rangle\right)=0\,.
\end{equation}
Then, \reg{vop} gives
\begin{eqnarray}
&&\sum_{n=2}^\infty{\lambda_n\over ((n-1)!)^2}\sum_p
(p+1)(p+2)\cdots(p+n-1)\\
&&\hskip30pt\cdot(-p+1)(-p+2)\cdots(-p+n-1) :a_pa_{-p}:
+
\lambda_1\sum_p :a_pa_{-p}: = 0\,.
\end{eqnarray}
Considering the $p=1$ terms gives us $\lambda_1=0$, while the terms
for $p>1$ give
\begin{equation}
\lambda_1+\sum_{n=2}^p\lambda_n(-1)^{n-1}\left({p+n-1\atop n-1}\right)
\left({p-1\atop n-1}\right)\,,
\end{equation}
and so we deduce that $\lambda_n=0$ for all $n$ as required.

Note that the trick in the choice of basis states is to choose all the
oscillator levels to go off to infinity as the level of the state
does. If we had chosen, as in the above, at least one $a_{-1}$, then
every $\lambda_n$ would have been involved in every term $a_pa_{-p}$,
leading to an uncontrollable infinite set of equations. The same idea however
cannot be applied to states involving more than two oscillators.

For the states $a_{-1}a_{-n}a_{-n}|0\rangle$, the above
technique fails, as arbitrarily large conformal weight states
contribute to all the operators. So, to prove that the method cannot
generalize beyond two oscillators, we only have to show that
$a_{-1}a_{-n}a_{-n}|0\rangle$ is not an $L_{-1}$ descendant (this will
also be part of the required proof to show that the above set of
states is a suitable basis if we wished to proceed with the method).
The proof is by contradiction, exactly as in (\ref{uno}-\ref{tre}).

Let us instead take a more general approach, though this reveals less about
the explicit structures involved, and consider a
quasi-primary state $\psi$ such that $V_{-1}(\psi)=0$. We wish to show
that $\psi=\lambda|0\rangle$ for some $\lambda\in\ce$.
Now, since $\psi$ is quasi-primary, \cite{PGmer}
\begin{equation}
[L_n,V(\psi,z)]=z^n\left(z{d\over dz}+(n+1)L_0\right)V(\psi,z)\,,
\end{equation}
for $n=\pm 1$,
giving
\begin{equation}
[L_1,V_{-1}(\psi)]=V_0(L_0\psi)
\end{equation}
and
\begin{equation}
[L_{-1},V_0(L_0\psi,z)]=-V_{-1}((L_0-1)L_0\psi)\,.
\end{equation}
Since $(L_0-1)L_0\psi$ is still quasi-primary, we may repeat the procedure,
and find that
\begin{equation}
V_0\left(\left((L_0-1)L_0\right)^nL_0\psi\right)=V_{-1}\left(\left(
(L_0-1)L_0\right)^n\psi\right)=0\,,
\end{equation}
for all $n\geq0$. For $\psi=\sum_n\psi_n$ where $L_0\psi_n=n\psi_n$,
this gives $V_0(\psi_n)=V_{-1}(\psi_n)=0$ for $n>1$.

This now reduces
the problem at each level to a finite one, and hence should render it
tractable.
For example, the first non-trivial level for the product of three
creation oscillators in the vacuum sector, as we discussed above, is
at conformal weight 5. A general state can be written as
\begin{equation}
\psi_5=\lambda_{113}a_{-1}a_{-1}a_{-3}|0\rangle+\lambda_{122}
a_{-1}a_{-2}a_{-2}|0\rangle\,.
\end{equation}
Now
\begin{equation}
V(\psi_5)_0={\lambda_{113}\over 2}\sum_{p+q+r=0}(r+1)(r+2)
:a_pa_qa_r:+\lambda_{122}\sum_{p+q+r=0}(q+1)(r+1):a_pa_qa_r:\,,
\end{equation}
and so we see, from considering the terms in $a_{-1}a_{-1}a_{2}$, that
$\lambda_{113}=0$ (and hence also that $\lambda_{122}=0$). Thus,
$\psi_5=0$ as required.

In general, consider a state in the vacuum sector containing $n$
creation operators, say
\begin{equation}
\psi=\sum_{\{ij\ldots k\}}\lambda_{ij\ldots k}a_{-i}a_{-j}\cdots
a_{-k}|0\rangle\,,
\end{equation}
where we have ordered $i\leq j\leq\cdots\leq k$. We say that
$\lambda_{ij\ldots k}>\lambda_{lm\ldots n}$ if $i>l$ or ($i=l$ and
$j>m$), and so on. Then we take the $\lambda$'s in ascending
order, and for
$\lambda_{ij\ldots kl}$ consider the term of $V_0(\psi)$ in
$a_{-i}a_{-j}\cdots a_{-k}a_{i+j+\cdots+k}$.

Now, we know from the form of the vertex operators that the
contribution from the creation operator $a_{-i}$ in $\psi$ to $a_r$
vanishes if $1\leq r\leq i-1$, and we see that we can deduce in turn
that all the $\lambda_{ij\ldots k}$, and hence $\psi$ itself, must
vanish, as required.

For $\psi=\prod a|\lambda\rangle$ (using an obvious schematic notation),
$V(\psi)_n$ is
\begin{equation}
\label{trifle}
\sum_mV_m\left(\prod_{a\perp\lambda}a|0\rangle\right)V_{n-m}(\hbox{remainder})\,,
\end{equation}
and we reduce our considerations to one-dimension as above ({\em i.e.}
either $V_0\left(\prod_{a\perp\lambda}a|0\rangle\right)=0$ or
$V_{0}(\hbox{remainder})=0$).

[Note that we consistently are using the fact that
$V$(subspace)$\subset$subspace (of the vector space spanned by $\Lambda$), and
that
operators associated with these subspaces commute (normal
ordering of \reg{trifle} is unnecessary).]

Now, suppose that $V_0(\phi|\lambda\rangle)=0$ for $\phi$ a
combination of creation operators (in the direction of $\lambda$) of
weight $N$. We wish to show that $\phi=0$.
Consider the form of the vertex operator. The $0$'th mode acting
in the sector with momentum $r\lambda$ is given by
\begin{equation}
\sum_{m\geq 0\atop p\geq 0}Y_mV_{n+m-p}(\phi)X_p\,,
\end{equation}
where $n=(r+{1\over 2})\lambda^2$ and $X_p$ and $Y_m$ are appropriate
modes of
$e^{\mp\lambda\cdot\sum_{s>0}{a_{\pm s}\over s}}$.

Consider first the term involving a single oscillator ($a_n$). This is
\begin{equation}
\label{trifling}
\left(V_n(\phi)|_{a_0=r\lambda}\right)_{\hbox{single-oscillator}}+
(-1)^N{\lambda\cdot a_n\over n}\sum_m(-r\lambda)^m\,,
\end{equation}
where the sum over $m$ is over the terms in $\phi$ formed from a
product of $m$ creation operators. (Note that the $a_0$ term in
$V(a_{-t}|0\rangle,z)$ is $z^{-t}(-1)^{t-1}a_0$.)

Naively, we could vary $r$ and effectively get vanishing of
arbitrarily large modes of the operator part of the vertex operator, which
should give vanishing of the state. However, the contribution to terms with
lower numbers of oscillators from powers of $a_0$ occurring in higher
states is difficult to handle, and in any case
we can instead use a very simple argument. Consider just the
case $r=0$. Then
\begin{equation}
\left(V_{{1\over 2}\lambda^2}(\phi)|_{a_0=0}\right)_{\hbox{single-oscillator}}
=0\,,
\end{equation}
and we deduce that the single-oscillator term in $\phi$ ({\em i.e.} $a_{-N}$)
vanishes. Then we consider the two-oscillator terms. There is no contribution
from $X$ and $Y$ (other than 1!), since the single-oscillator term in $\phi$
vanishes. The result for the zero-momentum sector which we have already proven
tells us that these vanish (the vanishing of $V_{{1\over 2}\lambda^2}(\phi)$
implies, from the $L_{-1}$ commutation relation, that
$V_0\left(L_0(L_0+1)\cdots(L_0+{1\over 2}\lambda^2-1)\phi\right)$ vanishes).
Proceeding in this way, we deduce that $\phi=0$,
as required.
\section{Determinism of the theories $\widetilde\Hil(\Lambda)$}
\label{twisted}
We must show that vanishing of
\begin{equation}
{\cal V}\left((\psi,\chi)\right)_0=\left(
\begin{array}{cc}V(\psi)_0 & \overline W(\chi)_0 \\
W(\chi)_0 & V_T(\psi)_0
\end{array}\right)
\end{equation}
implies vanishing of $\psi$ and $\chi$ (under suitable restrictions on the
states).
For the state $\psi$, the results of the previous section are sufficient (note
that
we do not even have to consider the structure of the twisted vertex operators
$V_T$). However, the twisted structure enters into consideration of the state
$\chi$.

It should be noted that it is not possible to reduce to a one-dimensional
problem
as we did in the previous section. Consider, for example,
\begin{equation}
\label{ope}
\overline W(c^i_{-{1\over 2}}\chi,z)c^j_{-{1\over 2}}\chi\,,
\end{equation}
where $i\neq j$ and $\chi$ is in the spinor ground state.
If this contains negative powers of $z$, then the operator product expansion
\reg{ope1} of
${\cal V}(c^i_{-{1\over 2}}\chi,z)$ and ${\cal V}(c^j_{-{1\over 2}}\chi,w)$
has singular terms and so leads to a non-zero commutator
on contour integration \cite{Ginsparg}, unlike in the untwisted case,
and we can no longer treat the orthogonal (vectorial) subspaces
independently.
Now, \reg{ope} is
\begin{equation}
\sum_n\overline W_n(c^i_{-{1\over 2}}\chi)z^{-n-{1\over 2}-{d\over 16}}
c^j_{-{1\over 2}}\chi.
\end{equation}
Since the state resulting from the action of the $n$'th mode has weight
${d\over 16}+{1\over 2}-n$, then for negative powers of $z$ we simply
require a non-zero piece in \reg{ope} of weight $\leq{d\over 8}$. Putting
in the explicit expression for $\overline W$, we see this is clearly true,
even in the case $d=8$ (at least if there exist vectors of length squared
two in the lattice). Thus, the twisted sector mixes up the
independent dimensions of the untwisted sector, and is implicitly tied up
with the momenta ({\em c.f.} $V(a^i_{-1}|0\rangle,z)$ and $V(|\lambda
\rangle,w)$ do not commute for $\lambda^i\neq 0$).

Take $\chi$ to be of conformal weight $N$, as we did for the untwisted
state in the previous section.
Now, using the explicit results of \cite{DGMtwisted},
\begin{eqnarray}
\overline W(\overline\chi,z)\chi&=&W\left(e^{z^\ast L_1}{z^\ast}^{-2L_0}\chi,
1/z^\ast\right)^\dagger\chi\\
&=&\sum_{\lambda\in\Lambda}\langle\chi|{\gamma_\lambda}^\dagger
z^{-2L_0}e^{zL_{-1}}:e^{B(1/z^\ast)^\dagger}:e^{{1\over z}L_1}|\chi\rangle
e^{A(1/z^\ast)^\dagger}|\lambda\rangle\,,
\end{eqnarray}
where $A(w)$ and $B(w)$ are of the form
$A(w)=A p^2 \log w+\sum_{n,m\geq 0}A_{mn}a_m\cdot a_n w^{-n-m}$
and $B(w)=\sum_{n\geq 0,s}B_{ns}a_n\cdot c_s w^{-n-s}$.
But $\overline W(\overline\chi)_0\chi$ is the weight $N$ piece of this, and so
if we consider a term in the sector with momentum $\lambda$ for $\lambda^2
=2N$ (we must check that such an assignment is possible -- see below
for a discussion), then we must have vanishing of
\begin{equation}
\langle\chi|{\gamma_\lambda}^\dagger e^{L_{-1}}e^{B^\dagger}e^Be^{L_1}
|\chi\rangle\,,
\end{equation}
where
\begin{equation}
B=-\sum_{s>0\atop{s\in\ze+{1\over 2}}}{\lambda\cdot c_s\over s}\,.
\end{equation}
Adjust the ground state spinor, if necessary, of one of the $\chi$ states
so that the ground state matrix element is non-zero ({\em e.g.} multiply it
by $\gamma_\lambda$), and then we see that
\begin{equation}
||e^Be^{L_1}
|\chi\rangle||=0\,,
\end{equation}
and hence that $\chi=0$, as required.

Now we consider whether it is possible to choose $\lambda\in\Lambda$ such
that $\lambda^2=2N$ for any $N\in\ze_+$. We have the restriction that $\sqrt 2
\Lambda^\ast$ be even (so that the twisted conformal field theory is
consistent), but even self-duality is clearly not sufficient. For example, of
the 24
even self-dual lattices in 24 dimensions, it is possible to make such a choice
for all but one of the lattices, namely the Leech lattice. As is well known,
the
Leech lattice has no vectors of length squared two. But this corresponds to
conformal weight one, and the lowest weight in the twisted sector after
projection is two, so there is in fact no problem in this case.

The general situation remains to be understood. However, note that the
minimal norm for an even self-dual (Type II) lattice \cite{ConSlo} is
$2\left[{d\over 24}\right]+2$, so that if the coefficients in the theta
series of the lattice are all non-zero above the minimal norm then we
have the required result. In any particular case, this is trivial to check,
and even if the result does not hold, we can always choose a $\lambda$ such
that
$\lambda^2$ lies within the minimal norm of $2N$, and we have to simply extend
the
above analysis by considering the appropriate number of terms in the expansion
of
$e^{A(1/z^\ast)^\dagger}$.
\section{Conclusions}
We have proposed a condition which will guarantee that any (inner) continuous
automorphism of a bosonic meromorphic hermitian conformal field theory
is generated by the zero modes of the vertex operators corresponding to the
states of conformal weight one, and checked that this is a reasonable
condition to impose on a conformal field theory by showing that it holds
for the FKS theories $\Hil(\Lambda)$. The proof that the twisted theories
$\widetilde\Hil(\Lambda)$ are deterministic depends on the theta function of
the
lattice being suitably behaved, though it is clear that a (more intricate)
proof not relying on such assumptions can be found. In any case, the proof
holds for the known self-dual theories at central charge 24 or less, and
in particular is consistent with Frenkel, Lepowsky and Meurman's result
\cite{FLMbook}
that there
are no continuous automorphisms of the Monster conformal field theory.
\label{conclusions}
\appendix
\section{Equivalence of definitions of determinism}
\label{app2}
Suppose $\Hil$ is such that $V_0(\phi)=0$ and $\phi$ quasi-primary implies that
$\phi$ is of weight
one (the same statement as that in the text by linearity of the vertex
operators in their arguments).
Then suppose that $V(\psi)_{-1}=0$. If $\psi$ is quasi-primary,
the usual commutation relation with $L_1$ \cite{PGmer} gives $V_0(L_0\psi)=0$,
and
so $L_0\psi$, by the assumption, is of weight one (clearly being
quasi-primary). Hence
$\psi=\lambda|0\rangle+\chi$, for some $\lambda\in\ce$ and $\chi$ of conformal
weight one.
But $V_{-1}(\lambda|0\rangle+\chi)|0\rangle=\chi$, and so $V_{-1}(\psi)=0$
implies $\chi=0$, and we have
the required result.

Conversely, suppose that $V_{-1}(\psi)=0$ and $\psi$ quasi-primary implies that
$\psi=\lambda|0\rangle$ for some $\lambda\in\ce$. Then, if $V_0(\phi)=0$,
$[L_{-1},V_0(\phi)]=0$, {\em i.e.} $V((L_0-1)\phi)_{-1}
=0$. If $\phi$ is quasi-primary, then so is $(L_0-1)\phi$, and so we have
$(L_0-1)\phi=\lambda|0\rangle$ for some $\lambda\in\ce$, {\em i.e.}
$\phi=-\lambda|0\rangle+\chi$, for some state $\chi$ of weight one.
But $V_0(\phi)=0$, and so $\lambda=0$. Thus, we have the required
equivalence.
\vfill
\eject

\end{document}